\documentclass[12pt,preprint]{aastex}

\newcommand{\et}{et al.}
\newcommand{\kms}{km s$^{-1}$}
\newcommand{\ha}{H$\alpha$}
\newcommand{\hb}{H$\beta$}
\newcommand{\solar}{\ifmmode_{\sun}\;\else$_{\sun}\;$\fi}

\newcommand{\HII}{H$\,${\sc ii}}

\newcommand{\vhelio}{V$\rm _{helio}$}
\newcommand{\coldens}{cm$^{-2}$}

\begin{document}

\title{The Stellar and Gas Kinematics of Several Irregular Galaxies}

\author{Deidre A.\ Hunter\footnotemark[1]}
\affil{Lowell Observatory, 1400 West Mars Hill Road, Flagstaff, Arizona 86001
USA}
\email{dah@lowell.edu}

\author{Vera C.\ Rubin\footnotemark[1]} 
\affil{Carnegie Institution of Washington, 5241 Broad Branch Road, NW, 
Washington, D.\ C.\ 20015 USA}
\email{rubin@dtm.ciw.edu}

\author{Rob A.\ Swaters}
\affil{
Department of Physics and Astronomy,
Johns Hopkins University, 3400 N. Charles Street and Space Telescope
Science Institute, 3700 San Martin Dr.,
Baltimore, MD 21218}
\email{swaters@pha.jhu.edu}

\author{Linda S.\ Sparke}
\affil{Washburn Observatory, 475 North Charter Street, 
Madison, WI 53706-1582 USA}
\email{sparke@astro.wisc.edu}

\and

\author{Stephen E.\ Levine}
\affil{US Naval Observatory, Flagstaff Station, PO Box 1149, 
Flagstaff, AZ 86002-1149 USA}
\email{sel@nofs.navy.mil}

\footnotetext[1]{\rm Visiting Astronomer, Kitt Peak
National Observatory, National Optical Astronomy Observatory, which
is operated by the Association of Universities for Research in Astronomy, Inc.
(AURA) under cooperative agreement with the National Science Foundation.}

\begin{abstract}

We present long-slit spectra of three irregular galaxies from which we
determine the stellar kinematics in two of the galaxies (NGC 1156 and
NGC 4449) and ionized-gas kinematics in all three (including NGC 2366).
We compare this to the optical morphology and to
the HI kinematics of the galaxies.
In the ionized gas, we see a linear velocity gradient 
in all three galaxies. 

In NGC 1156 we also
detect a weak linear velocity gradient in the stars
of (5$\pm$1/$\sin i$) \kms\ kpc$^{-1}$ to a radius of 1.6 kpc.
The stars and gas are rotating about the same axis, but this
is different from the major axis of the stellar bar which dominates
the optical light of the galaxy.

In NGC 4449 we do not detect organized rotation of the stars and
place an upper limit of (3/$\sin i$) \kms\ kpc$^{-1}$ to a radius
of 1.2 kpc.
For NGC 4449, which has signs of a past interaction with another galaxy,
we develop a model to fit the observed kinematics of the stars and gas.
In this model the stellar component is in a rotating disk
seen nearly face-on while the
gas is in a tilted disk with orbits whose planes 
precess in the gravitational
potential. This model reproduces the apparent counter-rotation of the inner
gas of the galaxy. The peculiar orbits of the gas are presumed due
to acquisition of gas in the past interaction.

\end{abstract}

\keywords{galaxies: irregular --- galaxies: kinematics and dynamics 
--- galaxies: structure
--- galaxies: individual (NGC 1156, NGC 2366, NGC 4449)}

\section{Introduction}

We have learned that irregular galaxies are not just small versions
of spiral disks, but we do not know the intrinsic
characteristics of these systems,
the most common type of galaxy in the universe.
Indeed, the fundamental structure of irregular galaxies is still in 
doubt. It has been generally believed that irregular galaxies are
disks like spirals, and like spirals,
surface brightness profiles in most irregulars are most often adequately
fit with an exponential law (Patterson \& Thuan 1996; van Zee 2000).
However, studies of the distributions of projected minor-to-major axis ratios
$b/a$ suggest that irregulars are thicker than spirals, perhaps having
intrinsic flattening ratios $(b/a)_0$ of 
0.3--0.4 rather than the 0.2 value generally adopted for spirals
(Hodge \& Hitchcock 1966, van den Bergh 1988, Binggeli \& Popescu 1995).
Staveley-Smith, Davies, \& Kinman (1992), on the other hand,
argue that the low ratio of
rotation velocity to velocity dispersion often seen in irregulars
must imply a thick disk with an intrinsic flattening
ratio as high as 0.6. 
More recently Sung \et\ (1998) have also analyzed 
ellipticity distributions and concluded
that dwarf irregulars and Blue Compact Dwarfs (BCDs) 
are more triaxial than disk-shaped, 
having axis ratios 1:0.7:0.5 and 
being only a little less spherical than dwarf ellipticals. 
What then is the true shape of irregulars?

Because of these outstanding issues, we have begun a program aimed at
understanding
the intrinsic structure of normal irregular galaxies
through determining their stellar kinematics.
In this paper we present results from long-slit spectral observations
of stellar absorption features in two 
irregular galaxies of high surface brightness.
While observations of giant galaxies are often detailed enough
to justify sophisticated analysis of their structure, 
(see, for example, Statler \& Fry
1994; Statler, Smecker-Hane, \& Cecil 1996; Statler \& Smecker-Hane 1999),
observations of the stellar kinematics 
of irregular galaxies have been harder to obtain
(but see Swaters 1999).
Most irregular galaxies are much lower in surface brightness
than giant galaxies, and they rotate much more slowly,
so a higher velocity resolution is needed.
Thus, the signal-to-noise that is possible with
current instrumentation limits what can be deduced for irregular galaxies.
Nevertheless, we can begin to learn about the highest surface brightness
irregulars now, and expect to extend to lower surface brightness objects
as long-slit spectrographs or field integrators with moderate spectral resolution
become available on very large telescopes.

An understanding of the shape of Im galaxies is important to understanding
other aspects of these systems.
A colleague of ours saw a broad-band optical
picture of the irregular galaxy Sextans A
(Hunter 1997)
on a bulletin board and asked why the galaxy appears square
and has such sharp edges. Sextans A is not alone; other
irregular galaxies also defy our expectations of what galaxies should look like. 
The rectangular shape of these galaxies is often taken to mean that the
galaxies are barred. But, if that is the case, 
their bar structures are very different
from those in spirals. In spirals the bar length 
is typically $\leq$0.3 of the size of the stellar disk
(Elmegreen \& Elmegreen 1985). In the rectangular
irregulars the
bar is a significant fraction of the entire optical galaxy. 
For example, in
NGC 4449 the bar has a length that is 90\% of D$_{25}$, the diameter measured
to a B-band surface brightness level of 25 magnitudes arcsec$^{-2}$
(Hunter, van Woerden, \& Gallagher 1999). 
What does it mean for most of the galaxy to be a bar?

There are also often peculiarities in the gas kinematics and distributions
that are not evident in the optical morphology.
In Sextans A the velocity field of the neutral gas is not aligned with
the optical axis of symmetry and there is some evidence that gas is moving
on elliptical orbits (Skillman \et\ 1988). In NGC 4449 the inner gas is
counter-rotating with respect to the outer gas and neither gas system is
aligned with the isophotes of the 
optical galaxy (Hunter \et\ 1999). However, because of
its large cross-section, the gas of
many irregulars could have been perturbed by outside forces
over the galaxy's lifetime (for example, NGC 4449: Hunter \et\ 1998).
What are the kinematics of the stellar system, which has a smaller 
cross-section,
and how do they compare to that of the gas?

In this paper we report results of our first observations
of stellar kinematics determined from absorption spectra in 
the irregulars NGC 1156 and NGC 4449. We also discuss the kinematics
of the ionized gas in these two galaxies as well as in NGC 2366.

\section{The Galaxy Sample}

Two of the galaxies in our sample are high surface brightness 
irregular galaxies: NGC 1156 and NGC 4449.
They have $\mu_{25}$, the average surface brightness
within D$_{25}$, of 22.4 and 22.1, respectively.
The third galaxy, NGC 2366, is a more typical, lower surface
brightness irregular ($\mu_{25}\sim23.4$). 
All three galaxies are classified as
IBm, barred Magellanic-type irregulars, 
by de Vaucouleurs \et\ (1991; $\equiv$RC3).
Only NGC 2366 might be considered a ``dwarf'' irregular under any of the
more common luminosity-based definitions adopted by authors.
In Table \ref{tabgal} we list some global properties of the galaxies
that are pertinent to our discussion.
Figure \ref{figv} shows V-band and \ha\ images of each galaxy.

At an M$_B$ of $-$18.0,
NGC 1156 is at the high end of the range of luminosities seen in normal
irregular galaxies (Hunter 1997), and is about 25\% brighter
than the Large Magellanic Cloud in the B-band.
NGC 1156 has a modestly high rate of star formation
and \HII\ regions are crowded over the disk.
Karachentsev, Musella, \& Grimaldi
(1996) describe NGC 1156 as ``the less [sic] disturbed galaxy in the Local
Universe,'' and there are no significant galaxy neighbors within 0.7 Mpc
and $\pm$500 \kms. 
However, in HI channel maps McIntyre (2002) sees a structure
that resembles a tiny tidal arm. It terminates at the northeast
HI complex, where there is an unusually large range in velocity
(Swaters 1999), and extends 3\arcmin\ to the southeast. Although no
other evidence of companions is seen, it is possible that NGC 1156
is not the pristine galaxy that it once seemed.

NGC 4449 is one of the most luminous and actively star-forming
IBm galaxies
(RC3, Hunter 1997).
It has an integrated M$_B$ of $-$18.2, making it 1.4 times
as luminous as the LMC. It is 
forming stars twice as rapidly as the LMC
even though the sizes of these two galaxies
are similar.
NGC 4449 is also unusual in having neutral hydrogen gas
at 10$^{19}$ atoms cm$^{-2}$ 
extending to six times its Holmberg radius 
(van Woerden, Bosma, \& Mebold 1975;
Bajaja, Huchtmeier, \& Klein 1994),
a distribution that is about three
times more extended than that of
most other irregular galaxies (Huchtmeier,
Seiradakis, \& Materne 1981).
Van Woerden \et\ (1975) 
and Bajaja \et\ (1994)
found opposite velocity
gradients between gas in the inner few arcminutes and the gas beyond that.
Hunter \et\ (1998) 
resolved part of the extended gas into
enormous streamers that wrap around the galaxy.
These, and the counter-rotating gas systems,
suggest that NGC 4449 has undergone an interaction and possibly 
a merger with
another galaxy (Hunter \et\ 1998).
NGC 4449 is a member of the CVnI ``loose'' cloud of galaxies
(de Vaucouleurs 1975), and the small irregular DDO 125 is
only 41 kpc away in the plane of the sky. Models by Theis \& Kohle
(2001) suggest that DDO 125 was the interaction partner that produced
the streamers of gas that we now see around NGC 4449, but their
models do not address the counter-rotating gas system in the central
part of the galaxy.

NGC 2366, located in the M81 Group of galaxies,
is a lower luminosity irregular, having almost the same absolute
B magnitude as the
Small Magellanic Cloud. It contains the supergiant \HII\ complex NGC 2363
to the southwest of the galaxy center. NGC 2363
is nearly twice as bright in \ha\ as the 30 Doradus nebula in the
Large Magellanic Cloud and contains a large fraction of the total
current star-formation activity of the galaxy (Aparicio \et\ 1995,
Drissen \et\ 2000).
However, there is also another large \HII\ complex to the west of 
NGC 2363 and
numerous smaller \HII\ regions scattered along
the disk.

\section{The Spectroscopic Observations, Data Reduction, and Analysis}

\subsection{Long-slit spectra}

We obtained long-slit spectra of our three galaxies during a 4-night
observing run in 2000 January
at the Kitt Peak National Observatory (KPNO) 4 m telescope.
We used the RC-spectrograph and a Tektronics 2048$\times2048$
CCD detector.  The detector gain was 3.1 e$^{-}$ ADU$^{-1}$, and the readnoise
was 4.0 e$^-$.
We used a 3\arcsec-wide slit without a decker in order
to obtain the longest slit possible.
The useful portion of the slit was 463 pixels (5.5\arcmin) in length, and from 
a series of star observations we determined the scale to be 0.71\arcsec\
pixel$^{-1}$. The stellar FWHM (along the slit) 
was $\sim$2.5\arcsec\ near the center 
of the spectra.
The focus degraded both blueward and redward of that so that
at 4840 \AA\ the FWHM of a star was 6\arcsec\ and the profile became
double-peaked with a central minimum; at 5274 \AA\ the FWHM
was 3.1\arcsec.

With grating BL380 in second order and order-separating
filter BG39 we covered 4800--5350 \AA\
at a dispersion of 0.36 \AA\ pixel$^{-1}$. The spectral resolution was 
2.0--2.5 pixels, measured from comparison lamp emission lines and night
sky lines. A typical value of 2.2 pixels
corresponds to 47 \kms\ at [OIII]$\lambda$5007.
We chose this particular wavelength region in order to include
the MgIb absorption features at
5183.6 \AA, 5172.7 \AA, and 5167.3 \AA.  These features are stronger in the
cooler stars (late F through M).
However, the cross-correlation technique 
includes all absorption lines detected in the spectra, not just
the MgIb features. In addition to the absorption lines, our spectra
included emission lines, the strongest of which are
\hb\ and [OIII]$\lambda$4959, 5007.
These emission lines allow us to obtain the radial velocities of the 
ionized gas for comparison to that of the stars.

We used an FeAr comparison lamp for wavelength calibration, and took a
calibration observation before
and after each object spectrum.
The electronic pedestal was subtracted using the overscan strip of each
image and separate bias frames to remove two-dimensional structure.
Pixel-to-pixel variations were removed using observations of a white
screen hanging on the inside of the dome. We also obtained twilight sky
flats to determine the slit illumination function, but this was not
an important correction for our purposes and it was applied only to the
version of the spectra with linear wavelength repixelization.
We observed a star stepped along the slit in order to remove S-distortion,
but found that the amount of distortion was small compared to errors
in fitting the centers of the stellar spectra and so did not apply this
correction.
During the run we observed radial velocity standard stars 
from the United States Nautical Almanac Office (2000) and other
late-type stars chosen from Jacoby, Hunter, \& Christian (1984) 
to use as templates in the cross-correlation. The template stars that we used
are listed in Table \ref{tabstars}. We also used the twilight sky
as the template of a G2V star.
There was thin cirrus during a few observations, especially on the first night,
but skies were clear otherwise.

We observed NGC 1156 and NGC 4449 at a different position angle (PA)
each of the four
nights of the observing run. We oriented the slit along
the optical major
axis determined from a deep V-band image, 
along the minor axis, and at plus and minus 45\arcdeg\ from the major axis.
{\it Note that we define the position along the slit 
measured from the center of the galaxy in the direction
of the given position angle as positive, and positions along the slit
in the opposite direction as negative.}
Each slit was placed on the galaxy using a center predetermined
from the V-band images, as given in Table \ref{tabgal}.
We offset to these positions from nearby stars.
One exception was NGC 1156's spectrum at
PA 174\arcdeg\ which was offset 2\arcsec\ east of the original
position in order to avoid
a bright star cluster.
At each PA we observed a series of 2700 s exposures: 5--6
each of NGC 1156 and 4 each of NGC 4449. The object spectra were
preceded, interleaved, and followed with comparison lamp observations. 
Because NGC 4449 is so large,
observations of it were also preceded, interleaved, and followed with
separate sky observations taken 1\arcdeg\ (nights 1 and 2) or
6\arcmin\ (nights 3 and 4) alternately north and south of the galaxy.

Because NGC 2366 is much lower in surface brightness than NGC 1156 and
NGC 4449, we did not expect to detect absorption features well enough
to measure stellar kinematics in this galaxy in the available time.
Therefore, we 
only observed this galaxy for the emission-line kinematics.
We obtained observations at 4 different PA: Three
were placed on the center of the galaxy determined
from our V-band image; the fourth at PA 27\arcdeg\
was centered at 7$\rm ^h$ 29$\rm ^m$ 02.5$\rm ^s$, 69\arcdeg\ 14\arcmin\
17\arcsec\ (2000) in order to reach the nebulae in the outer part of the galaxy.
Exposures at PAs 30\arcdeg\ and 120\arcdeg\
were 3600 s each;
spectra at PAs 63\arcdeg\ and 27\arcdeg\ were single 1800 s
exposures. A separate sky observation was obtained only for one of
the observations at
PA 30\arcdeg.

After the spectra were corrected for the electronic pedestal and pixel-to-pixel
variations, we did a two-dimensional fit to each of the comparison lamp
spectra.
We then performed a two-dimensional transformation of 
each object frame using the nearest comparison.
If an object frame was bracketed by comparisons, we did the
two-dimensional fit to
both comparisons simultaneously, effectively averaging any wavelength
shifts.
On one of the nights, moving a mirror into place for observations
of the comparison lamp generated an electronic signal that moved the grating
by two encoder units (6 pixels). The motion was discrete in time, and therefore,
the observations of the comparison lamp were readily used to remove this
shift.
Our transformation repixelizes in the dispersion direction in order
to impose a uniform wavelength scale, and it 
corrects for curvature along lines of constant
wavelength.
We did the transformations both to a linear wavelength scale and to
a logarithmic scale; the linear versions were used to measure the emission-line
radial velocities and the logarithmic versions were used to measure the
absorption velocities using cross-correlation techniques.

We checked that the final galaxy spectra were aligned
in wavelength by overplotting night sky lines. 
When multiple spectra were obtained for a given PA,
we combined them with a rejection algorithm to remove cosmic rays. 
For NGC 2366 where we had single exposures, we edited cosmic rays
by hand.
For NGC 1156, the background
was removed using the region of the slit beyond the galaxy.
The background from NGC 2366 spectra were removed in this way as well
for all but one of the two observations at PA
30\arcdeg\ for which a separate sky observation was available.
For NGC 4449, separate sky frames were combined and subtracted from
object frames.
For the stars, we fit the background on either side of the star profile.

\subsection{Cross-correlation and absorption velocities}

To determine the radial velocity of the stars from absorption lines,
we used a cross-correlation technique that matches
absorption lines of template stars with the object spectrum
(Simkin 1974; Tonry \& Davis 1979;
Statler 1994).
We began by extracting the one-dimensional
template star spectra from the two-dimensional spectra by summing
along the slit to where the starlight blended into the background.
For the object spectra, we wanted to increase the signal-to-noise
as much as possible but without spatially summing over 
regions corresponding to large changes in
radial velocity. Therefore, we extracted one-dimensional object
spectra by summing every
5, 15, and 25 rows (3.55\arcsec, 10.65\arcsec, 17.75\arcsec) and comparing 
the results for the different summing intervals. We finally adopted
the results for the 15-row sums, and those measurements are what are
presented here. 
Spectral regions with emission lines were excluded in the cross-correlation. 

Each object spectrum was compared 
with each F---K-type template star, giving an estimate of the 
heliocentric radial velocity for each fit.
We averaged these velocities, weighted by the 
formal uncertainties of the fit to the cross-correlation peak.
The uncertainties of each cross-correlation
are of order 5--20 \kms, depending on the signal-to-noise. 
There were no systematics greater than 0.3 \kms\ involving
any particular template star. 
On the one night that we did two radial velocity standards, cross-correlation
of one with the other reproduces the \vhelio\ to 2.5 \kms, and this
is probably a good measure of the uncertainties for high signal-to-noise
spectra and a lower limit on the uncertainty of the galaxy measurements.

DAH and RAS independently applied the cross-correlation technique to the
galaxy data, using different approaches to subtracting the continuum and
fitting the cross-correlation peaks. Comparing our results
gave us a good feeling for the real uncertainties in our results.
For all but a few points the agreement was within a few \kms, 
well within the formal
uncertainties. However, in NGC 1156 
the point at r$=-$42\arcsec\
in PA 129\arcdeg\ differed by 24 \kms,
the point at r$=-$52\arcsec\ in PA 84\arcdeg\ differed by 18 \kms,
and the point at r$=-21$\arcsec\ and $-31$\arcsec\ in PA 174\arcdeg\
differed by 11 \kms.
These differences are, nevertheless, within 1--2$\sigma$.
In addition to the two cross-correlation approaches, VCR
used the technique described below for measuring emission lines
to measure the centroid of the MgIb absorption lines in the central
parts of two of the NGC 1156 spectra. This is a completely different
approach from cross-correlation and involves inverting the spectrum
to make the absorption lines look like emission lines. Those measurements
agreed well with the results of the cross-correlation technique,
although it could only be applied to the brighter regions of the spectra.
All of these experiments gave us confidence in our absorption line
velocity measurements and their uncertainties.

We used the radial velocity standards as template stars
and cross-correlated
the absorption lines to determine the radial velocities of the other
template stars. To further fine tune the zero points, we determined
the central velocity V$_{sys}$ of the galaxy in each spectrum where that
was reliable to do and took the average. The radial velocities from
each spectrum were then adjusted to match this V$_{sys}$. The
measured and average V$_{sys}$ and amount $\Delta$V$_{sys}$ by which the
velocities measured from each spectrum 
were adjusted are given in Table \ref{tabvc}.

\subsection{Emission-line measurements}

The emission spectra were measured by VCR
using a customized measuring program in the software package VISTA
(Rubin, Hunter, \& Ford 1991). The program fits a 2-dimensional 
polynomial to lines across a spectrum, and then returns the wavelength and 
radial position for each pixel; no repixelling is done. 

For the current observations, several very minor modifications were made. 
First, the comparison FeAr frames for each PA were summed, using a 
weighting of 1 for a comparison taken between exposures, and a weight
of 0.5 for a 
comparison taken at the beginning or end of each set. This assured that the 
individual comparison frames were handled in the same manner as the galaxy
frames.
The (rectified, summed) comparison was then fit with a low order 
polynomial whose coefficients were transfered to the corresponding galaxy 
frame. For each of the 12 summed comparison frames, the fit to 21 lines
(4806 \AA --5328 \AA ), each sampled at 16 radial positions, 
had an rms of 0.04 \AA , 
equivalent to 2.4 \kms. We are satisfied that we have preserved the velocity 
and radial scales from the comparison arcs to this level. 
Even though preliminary tests had shown that 
velocities differed only by a few \kms\ in measuring the summed 
galaxy spectrum with any one of the bracketing comparisons, we preferred to sum 
the comparison frames so as to treat them in a manner analogous to that used 
for the galaxy spectra.

Only the measurements of the [OIII]$\lambda$5007 line are used in the analysis
that follows.
H$\beta$ emission is superposed on a broad absorption line, and 
[OIII]$\lambda$4959 is only one-third as strong as $\lambda$5007, causing their
measures to have lower weight.  
In addition, the focus of the spectra degrades below about 4950 \AA,
causing the emission lines to develop a faint red wing. 
We have confirmed that this does not affect the $\lambda$5007 emission line,
and that there is no correlation between line intensity and velocity.
We did measure both $\lambda$4959 and $\lambda$5007 for a few spectra
and found that they were the same to within a few \kms.
We adopt an intrinsic wavelength of 5006.853 \AA\ 
for [OIII] (Kaufman \& Sugar 1986).

The measuring program fits a Gaussian to the 
emission in a given line
at successive points along the slit, and the intensity-weighted
centroid of the 
measures is adopted as the velocity there. Tests have shown that the centroid is 
a more stable measure than the peak. In regions where the emission is strong, 
a measurement is taken at each pixel (0.71\arcsec) 
in the radial direction.
Where the emission is weak, measurements are integrated over 5 or 6 pixels
(3.5\arcsec--4.3\arcsec).
The accuracy of each radial velocity is about $\pm$5 \kms\ on the plane of 
the sky.

\subsection{Stellar velocity dispersion}

Fitting of the cross-correlations peaks also yields a FWHM of
the profile, which is a measure of the velocity dispersion.
The resolution of the cross-correlation fitting is of order 70 \kms\
(square-root of 2 times the spectral resolution).
For NGC 1156 the FWHM of the cross-correlation profiles are not resolved.
For NGC 4449, 
the profile may be marginally resolved, indicating a high stellar velocity
dispersion, but
the results depend strongly on the template star used.
The velocity dispersion in the neutral gas is also unusually high
in this galaxy: of order 20 \kms\ within the optical galaxy
rather than the usual 10 \kms\ seen in most gas disks (Hunter \et\
1999).
However, given the low resolution and the uncertainties resulting
from the choice of template star,
we do not feel confident in stating a
stellar velocity dispersion from these data for either galaxy.
The velocity dispersion has been measured in the 
Sm galaxy UGC 4325 ($=$NGC 2552) by Swaters (1999) and found
to be 19$\pm2$ \kms. A velocity dispersion this small is
not measureable with our data.

\section{Results: NGC 2366} \label{sec2366}

One spectrum was obtained with the slit placed at a PA of 30\arcdeg, 
along the major axis
as shown in Figure \ref{figv}.
One can see that this PA
fairly respresents the major axis of the galaxy at all
surface brightness levels (see Hunter, Elmegreen, \& van Woerden [2001] 
for more discussion).
Another spectrum was obtained with the slit placed along the 
morphological minor axis of 120\arcdeg, and
one at a PA of 63\arcdeg\ was designed to pick up the supergiant
HII region to the southwest. These three spectra were obtained
with the slits placed at the center of the galaxy.
The fourth spectrum at PA=27\arcdeg\ is displaced 
113\arcsec\ from the
center along PA=27\arcdeg\ in order to intersect the prominent nebulosity
to the north. 
The four slit positions are shown superposed on an \ha\ image
of the galaxy in Figure \ref{figv}.

Strong ionized-gas emission lines of H$\beta$
and [OIII]$\lambda$5007 and $\lambda$4959 are observed in all PAs.
Weak lines of Fe I (4986 \AA, 5015 \AA) are detected from the large 
southwestern emission region NGC 2363 and from
the northeastmost knot in PA=63\arcdeg.
In every case, these weaker lines arise in the regions with the strongest
emission at H$\beta$ and [OIII]$\lambda$4959,5007.
In PA=30\arcdeg\ only, an emission line is seen at 5096 \AA\
arising in a region  of relatively strong continuum but only moderate
line emission and extending into a region of low continuum. 
This line emission comes from the region northeast of the circular
ring, about 50\arcsec--67\arcsec\ from the adopted center.
In other knots with even stronger continua, this emission line is not observed.

We plot \vhelio\ measured in the plane of the sky
from each emission spectrum 
in Figure \ref{fighelion2366}. 
Linear rotation is observed generally northeast 
of the center in PA 27\arcdeg,
30\arcdeg, and 63\arcdeg\ to 1.9 kpc.
In PA=120\arcdeg, the morphological minor axis,
velocities are essentially flat, with peak-to-peak
velocity variations only of order $\pm$10 \kms. This would imply
a kinematical minor axis of PA=120\arcdeg, and major axis of about 30\arcdeg.

The ionized gas in NGC 2366 exhibits
deviations from overall
rotation due to the energetic output of massive stars
in star-forming regions. There
are two supergiant
\HII\ regions in the southwest part of the galaxy---NGC 2363 
and NGC 2366-III---that are known to
exhibit high kinetic energies in the ionized gas 
(Hunter 1982; Roy \et\ 1992; Hunter \& Gallagher 1997; Martin 1997).
We see this at negative RA at PAs of 30\arcdeg\ and 63\arcdeg\
in Figure \ref{fighelion2366}.

The ionized gas and HI have similar rotation velocities
in the inner part of the galaxy.
This can be seen in Figure \ref{fighin2366em} where the velocities
of the ionized gas
are superposed on position-velocity diagrams made from slices
through the HI data (Hunter \et\ 2001).
We see that the ionized gas velocities nearly follow the 
ridge-line of the neutral gas with mostly modest deviations of the ionized
gas from HI peaks.
Tomita \et\ (1998) found differences between 
the velocities of \ha\ and HI in NGC 2366 of order $\pm$10 \kms,
consistent with what we see. 

Major axis PAs have been measured from optical surface photometry
and from the HI velocity field as well.
From R-band surface photometry,
Swaters (1999) determined a PA of 28\arcdeg\ for the major axis
of NGC 2366.
Hunter \et\ (2001) used a V-band image
and found a PA of 32\arcdeg; the exponential fit deduced
from the V-band surface photometry to 7.5\arcmin\ radius
also fit well the J-band surface photometry as far as it was measured 
(4\arcmin\ radius).
These agree well with the nominal PA of 30\arcdeg\ for the emission-line
kinematics seen here.

Modeling of the HI velocity field, on the other hand, produces
different parameters.
The 21 cm observations do
indicate an overall rotation,
but with a notable asymmetry at large radii ($>$2\arcmin).
Modeling of 21 cm kinematic data over the galaxy as a whole
gives a major axis of 39\arcdeg--46\arcdeg\
(Braun 1995; Wevers, van der Kruit, \& Allen 1986; 
Swaters 1999; Hunter \et\ 2001).
The inclination angles infered from the HI velocity field modeling
are also different from those determined from the optical surface
photometry, being smaller by 7\arcdeg--14\arcdeg.
These differences suggest that the HI distribution or kinematics
deviate from axisymmetry; that is, there are noncircular motions in the
HI or the HI disk is warped.

We have examined several models to describe the optical emission-line kinematics
of the inner $\sim$2\arcmin\ radius:
1) major axis at PA=30\arcdeg\ (near that determined from the optical
surface photometry),
and 2) major axis at PA=45\arcdeg\ (near that determined from the
HI kinematics). Both models assume that the galaxy
is circular in its principal plane, that the adopted center is the center
of mass, and that internal motions
in the emission
regions are small with respect to the rotation velocities.  For a galaxy
with small rotation velocity, this latter assumption is likely to be
violated in
the strong emission complexes such as we see in the southwest portion
of NGC 2366.

{\it Model 1}: We adopt 30\arcdeg\ as the PA of the stellar
kinematical major axis.
A weighted least-squares fit to the 
line-of-sight velocities,
$+$110\arcsec $<$ r $<$ $-$46\arcsec, gives 
a gradient of (0.50$\pm$0.04/$\sin i$) \kms\ arcsec$^{-1}$
or (30.4$\pm$2.2/$\sin i$) \kms\ kpc$^{-1}$
(solid line in Fig.\ \ref{fighelion2366}). 
To the southwest at larger distances from the center,
the optical velocities are flat at about 80 \kms.
For an inclination of 72\arcdeg\ determined from optical surface photometry,
under the assumption that these are thick
disks with intrinsic $(b/a)_0$ of 0.3
(Hodge \& Hitchcock 1966; van den Bergh 1988),
the gradient in the plane of the galaxy is 
$32\pm$2 \kms\ kpc$^{-1}$.
This rotation curve, deprojected to PA=63\arcdeg\ and overplotted on the 
PA=63\arcdeg\ measured velocities (Fig.\ \ref{fighelion2366}), 
is a moderately good representation
of the observed velocities to the north; but not to those
in the south where non-circular motions dominate.
 
{\it Model 2}: For this model, we adopt the PA of the major axis as
$\approx$45\arcdeg, the approximate mid-point of 
27\arcdeg, 30\arcdeg\ and 63\arcdeg. Hence,
in all three PAs, velocities should be similar.
We show in Figure \ref{fign2366mod2}
line-of-sight velocities measured  in the
region $+$125\arcsec\ $<$ r $<$ $-$50\arcsec\ for these three PAs.
Their similarity over this region is striking. A weighted
least-squares fit produces a slope of
0.50$\pm$0.02 \kms\ arcsec$^{-1}$
(solid line in Fig.\ \ref{fign2366mod2}), 
the same as that for Model 1, above.
Along the
adopted major axis, PA$=$45\arcdeg, the rotation curve
would have a velocity gradient of ($0.52\pm0.02/\sin i$) \kms\ arcsec$^{-1}$
or ($31\pm1/\sin i$) \kms\ kpc$^{-1}$.
This rotation
curve,
projected onto the four PAs in the plane of the sky are shown in 
Figure \ref{fighelion2366} (dashed line); 
the radial extent of the rotation curve
deprojects to r $>-20$\arcsec\ along what is a near-minor axis
in PA$=$120\arcdeg\ in this model.
For $i=72$\arcdeg, the gradient is $33\pm1$ \kms\ kpc$^{-1}$.
 
There is little reason from these data 
for choosing between these models.
Both fit the northeast gradients along the three position angles.
Although the minor axis velocities
are reproduced by both models, the observed small 
velocity gradient at r$>-20$\arcsec\
in PA$=$120\arcdeg\ offers some evidence that a major axis near 
45\arcdeg\ is a slightly better fit.
For any model,
the lack of observed rotation for r$<-20$\arcsec\ on the southwest
side is assumed to be due to NGC 2363 and HII region NGC 2366-III
whose internal motions distort the
systemic velocity pattern. However, observations by
Odewahn (1989) suggest that there may also be a 
twisting of the isovelocity contours 
in this region that implies a more systematic deviation
from the velocity field in the rest of the galaxy.
 
The resulting rotation curves for NGC 2366 are shown in 
Figure \ref{figrot2366}. Superposed on this is the rotation curve
derived from the HI (Hunter \et\ 2001).
There is fair agreement between the optical and the
21 cm curves in the regions where they overlap.  
However, the nuclear rise of
the optical velocities is steeper
and the linear rotation extends further in the
ionized gas than in the HI. 
This is the result of
the differences in resolution of the two data; the FWHM of the
HI beam is 34\arcsec$\times$29\arcsec\ while the resolution of the
optical spectra is a few arcseconds.
Points at large distance from the center of the galaxy, however,
are consistent with the leveling-off in the rotation curve seen
in the HI.

\section{Results: NGC 1156} \label{sec1156}

A deep V-band image of NGC 1156 reveals a rectangular shape
that grows more circular at low surface brightness levels.
The logarithm of the V-band image is shown in Figure \ref{figv} 
in order to allow comparison of the morphology of the galaxy
at different surface brightness levels.
We had used this V-band
image to determine the morphological major axis of the galaxy,
39\arcdeg. A line is drawn along this axis in
Figure \ref{figv}.
From R-band surface photometry Swaters (1999) determined 
the major axis to be 37\arcdeg.
However, the HI velocity field
has a major axis position angle of 263\arcdeg\ ($=$83\arcdeg) overall,
but with some irregularities. In the region 
that we measure the \ha\ velocities, the
PA of the HI kinematics is of order 130\arcdeg.

In one of our observations the slit was placed along the
morphological major axis 39\arcdeg. Other spectra were obtained
along the morphological minor axis at 
129\arcdeg\ and $\pm$45\arcdeg\ from the major axis at
PAs of 84\arcdeg\ and 174\arcdeg. 
The four slit positions are shown superposed on our \ha\ image
in Figure \ref{figv}.
The center of the galaxy was
taken as the center of middle-level isophotes in the V-band image.

\subsection{Ionized gas}

We show \vhelio\ in the plane of the sky measured from [OIII]$\lambda$5007
for NGC 1156
in Figure \ref{fighelion1156em}. Superposed on any rotational trends
are large peculiar velocities in \vhelio.
The peculiar velocities are 
due to high kinetic energy in the ionized gas that resulted
from the energetic input of massive stars in HII regions.
NGC 1156 is actively forming stars and contains many 
concentrations of massive stars.
In our spectra we see deviations from rotation 
with line of sight velocities of order 12--20 \kms\ and extents along
the slit of order 300 pc to over 1000 pc.
These are consistent
with the expansion of giant and supergiant shells of ionized gas.
Fairly systematic undulations seen in PA 39\arcdeg\ repeat
in PA 174\arcdeg\ if the PA 174\arcdeg\ spectrum is flipped
so that spatially nearby regions are superposed. 
Thus, these position angles reveal little of the overall rotation,
and we have no alternative but to
determine the major axis of rotation from the remaining
spectra in PAs 84\arcdeg\ and 129\arcdeg.

In Figure \ref{fighelion1156em} we see evidence for ordered 
rotation
of the ionized gas with a linear velocity gradient, 
but not at the PA one might expect to see it.
For a rotating inclined axisymmetric disk, 
the steepest velocity gradient will be
along the apparent major axis. The morphological major axis is 39\arcdeg,
but rotation at this PA is not obvious.
We clearly see rotation in PAs 129\arcdeg, the morphological
minor axis, and PA 84\arcdeg. The velocities
in PA 174\arcdeg\ also appear relatively flat.

A weighted least squares fit to the points (excluding end points)
from r$=+$29.9\arcsec\
to r$=-36.6$\arcsec\ (1.6 kpc)
in PA$=$129\arcdeg\
is shown as the solid line in the panel of PA 129\arcdeg\ 
in Figure \ref{fighelion1156em}. 
Each of the spectra off the major axis should
exhibit velocities which are the projected component of the 
major axis velocities.  
The fit to PA 129\arcdeg\ is shown
transformed to the other PAs in the plane of the sky
under the assumption that 129\arcdeg\
is the kinematical major axis. 
These are the solid lines in the other panels in Figure \ref{fighelion1156em}.
A weighted least squares fit to the velocities in PA$=$84\arcdeg\
(excluding end points)
from r$=+$42.2\arcsec\ to
r$=-$19.2\arcsec\ 
is shown as the dashed line in panel PA$=$84\arcdeg\  in 
Figure \ref{fighelion1156em}.
This fit is also shown transformed to the other PAs under the assumption
that 84\arcdeg\ is the line of nodes.
These are the dashed lines in Figure \ref{fighelion1156em}.

We see that the velocity gradients in both PAs 84\arcdeg\ and 
129\arcdeg\ are quite similar.
Their similarity suggests that the line of nodes lies between
84\arcdeg\ and 129\arcdeg, with the resulting velocity gradient
being slightly steeper than the lines in Figure \ref{fighelion1156em}.
Thus, the ionized gas kinematical
major axis is not the same as the morphological major axis of the galaxy.
If the kinematical axis is 129\arcdeg, it is in fact the morphological 
{\it minor}
axis of the galaxy but agrees with the HI kinematics in the inner part
of the galaxy. However, if the ionized gas kinematical major axis
is 84\arcdeg, it is the same as the overall
HI kinematic axis.
Position angles are summarized in Table \ref{tabpa}.

The velocity gradients that we measure at the two PAs are fairly
similar. The fit to PA$=$129\arcdeg\ 
yields a gradient
of ($0.58\pm0.08/\sin i$) \kms\ arcsec$^{-1}$. This
becomes ($15.4\pm2.0/\sin i$) at our adopted distance.
The fit to PA$=$84\arcdeg\ 
yields a gradient
of ($0.51\pm0.04/\sin i$) \kms\ arcsec$^{-1}$.
This becomes ($13.5\pm1.1/\sin i$) at our adopted distance.
If the true kinematical major axis lies mid-way between
these two PAs, the velocity gradient would increase by a factor of 1.1
over the fit to PA 129\arcdeg\ and PA 84\arcdeg\ combined.

A comparison of the emission-line \vhelio\ with position-velocity
slices through the HI data of Swaters (1999) is shown in Figure
\ref{fighin1156em}. 
There is a minimum in the HI in the
center of the galaxy which is
surrounded by higher column density HI complexes.
Our ionized gas velocity measurements extend into the regions of
higher gas column density.
Given that there could be a small offset in V$_{sys}$ between the
optical and HI,
the agreement between HI and ionized gas velocities 
in PAs 84\arcdeg\ and 174\arcdeg\
is strong;
the ionized gas velocities generally fall along the peaks in the
neutral gas. However, at PA 129\arcdeg, the rotation of the 
ionized gas 
shows a steeper gradient than the HI. 

In PA 39\arcdeg\ the HI exhibits very broad velocity profiles at r$=+$75\arcsec.
This region is the HI complex to the northeast of the galaxy center
where the possible
tidal arm of McIntyre (2002) ends.
The optical velocities do not exhibit the
low velocities seen in the HI at this position. However,
our ionized gas velocities only just extend to this radius
and so we may be missing the region of lower velocities seen
in the HI.

\subsection{Stars}

The \vhelio\ measured from the stellar absorption lines
are plotted against position along the slit in Figure \ref{fighelion1156abs}.
In the figure we show the results from measuring along the
spectra in 10.65\arcsec\ sums.
This step size corresponds to intervals of 400 pc in NGC 1156.
The stellar velocities are compared to the HI and ionized gas
velocities in Figure \ref{fighin1156em}.

We see very weak rotation 
of the stars 
about the morphological 
major axis (39\arcdeg), morphological minor axis (129\arcdeg),
and mid-way between (84\arcdeg) to 1.6 kpc.
Rotation about 39\arcdeg\ and 129\arcdeg\ seems marginally stronger than
about 84\arcdeg, but rotation about these two axes implies that
neither can be the kinematical major axis since the other must
be the minor axis and show no rotation if the motion is in
an ordered disk.
The only cicular-orbit model that makes physical sense 
is one in which the kinematical
major axis is near PA 84\arcdeg.
In Figure \ref{fighelion1156abs} we show a least-squares fit
(excluding end points) of r$=$43.3\arcsec\ (1.6 kpc) to r$=-$41.9\arcsec\
in PA 84\arcdeg. We also show the consequences of assuming this PA is the
major axis and transforming it to the other PAs in the plane
of the sky.
These are the solid lines in the figure.
If 84\arcdeg\ is near the kinematical major axis, the minor axis
is 174\arcdeg, and the velocities in PA 174\arcdeg\ do appear
to be flat.

The ionized gas rotation (the dashed line in Figure \ref{fighelion1156em})
is shown as the dashed line in 
Figure \ref{fighelion1156abs}.
The gradient of rotation in the NGC 1156 stars is 
($0.18\pm0.06/\sin i$) \kms\ arcsec$^{-1}$ or 
($4.8\pm1.5/\sin i$) \kms\ kpc$^{-1}$.  
The slope is the same if we fit the points at PAs 39\arcdeg,
84\arcdeg, and 129\arcdeg\ together.
The velocity gradient in the stars is less than that in the ionized gas.
However, the uncertainties in the stellar rotation velocities are 
large enough to allow both a velocity gradient that is comparable to
that of the ionized gas and one in which there is no rotation of the stars.

In Figure \ref{figrot1156} we plot the rotation velocities
of the stars in the three PAs with rotation (that is, 
excluding PA 174\arcdeg), projected
onto the plane of the galaxy. 
We also include the
rotation curve of the ionized gas and that determined from the HI
velocity field (Swaters 1999) reflected about r$=$0 for comparison.
Note that the resolution of the HI is much lower (HPBW$=$30\arcsec)
than that of the optical spectra.

\section{Results: NGC 4449} \label{sec4449}

A deep V-band image of NGC 4449 has been analyzed by Hunter \et\ (1999).
They found that the PA of the major axis changes from the inner
galaxy to the outer galaxy.
In the inner galaxy ($<$2.2\arcmin) 
the major axis is at a PA of 46\arcdeg, and in the outer galaxy
($>$3.1\arcmin)
it is at 64\arcdeg. 
This twisting of the isophotes is apparent in Figure \ref{figv}
where we show the logarithm of the V-band image. A line is
drawn along the major axis of the inner galaxy at 46\arcdeg\ to guide the eye.
We obtained one spectrum with the slit placed along
this axis. 
Other spectra were obtained along
the minor axis of 136\arcdeg\ and $\pm$45\arcdeg\
from the major axis at 91\arcdeg\ and 1\arcdeg.
All four slit positions are shown superposed on the
\ha\ image in Figure \ref{figv}.


\subsection{Ionized Gas}

We plot \vhelio\ measured from [OIII]$\lambda$5007 in the
four spectra of NGC 4449
in Figure \ref{fighelion4449em}. 
As for NGC 1156, there are regions where the deviations 
from rotation in \vhelio\
are quite large. 
These peculiar velocities most likely are the result of the kinetic
energy of the ionized gas that has resulted from the
input from massive
stars in the star-forming regions in the galaxy.
NGC 4449 is well-known
for the signatures of this energy input, including many ionized gas shells
and filaments and broad \ha\ lines seen in high-dispersion spectra
(Hunter 1982; Hunter \& Gallagher 1990, 1992, 1997; 
Hunter, Hawley, \& Gallagher 1993; Martin 1997). 
In our spectra we see deviations from rotation that 
are consistent with expansion velocities of order 20---100 \kms\ and
extents along the slit up to 1200 pc.
These are similar to those that we saw in NGC 1156.
In addition we see emission along a single line of sight that is
double-valued, with a large velocity difference ($\Delta$V$\geq$100 \kms\
in PA 1\arcdeg\ at r$\sim-20$\arcsec, in PA 46\arcdeg\ at r$\sim$40\arcsec,
and PA 136\arcdeg\ at r$\sim-20$\arcsec). As in NGC 1156,
the velocity extremes that we detect in nearby positions may arise
from segments of the same structure.
Hunter \& Gallagher (1997) point out that the center of the galaxy is
the base of a 1.3 kpc diameter supershell and coincident with
X-ray emission.
Most of the regions of peculiar velocities along the slit coincide
with filaments or loops that are visible on the \ha\ image,
some of which are identified by Hunter \et\ (1993)
and marked in Figure \ref{fighelion4449em}. 

The input from massive stars in NGC 4449 is so energetic that it
complicates the determination of the rotation.
However, in PA$=$46\arcdeg\ we do see a velocity gradient that can
be fit with a straight line.
We fit the velocities, excluding the peculiar velocities,
from r$=$110.1\arcsec\ (2.1 kpc) to $-62.7$\arcsec, and found a gradient
of (0.22$\pm$0.01/$\sin i$) \kms\ arcsec$^{-1}$ (the solid line
in Figure \ref{fighelion4449em}).
For a distance of 3.9 Mpc, this gradient is
(11.8$\pm$0.7/$\sin i$) \kms\ kpc$^{-1}$.
To see if the rotation curve in PA$=$46\arcdeg\
is consistent with circular rotation in a planar circular disk, 
we have transformed it to the other PAs in the plane of
the sky (the solid lines
in Figure \ref{fighelion4449em}).
These projections fit the rotation data at the other PAs reasonably
well, with considerable deviations due to the energetics of the ionized gas.
Thus, the rotation of the ionized gas is consistent with 
a kinematic axis that is the same as
the morphological major axis.
Position angles are summarized in Table \ref{tabpa}.

The \vhelio\ of
the ionized gas is shown superposed on position-velocity slices through
the HI in Figure
\ref{fighin4449em}. 
The HI data are from
Hunter \et\ (1999), and we have extracted slices of the HI data cube
at each of the optical PAs, summing over one HI beamwidth.
We see that the ionized gas follows the
rotation of the neutral gas. In some cases the peculiar velocities seen
in the ionized gas also follow those in the HI; see, for example, 
PA 1\arcdeg\ at a radius of about $+$1\arcmin. In some cases the
ionized gas is not associated with peaks in the HI; see, for example,  
PA 91\arcdeg\ at an r of 0. But, in many cases,
the velocities of the ionized gas are close to or near HI peaks.
Typical deviations of the ionized gas from the velocities of the 
HI peaks is of order
15--20 \kms, with some as high as 50 \kms.
For the galaxy NGC 4214, which is very similar in luminosity to
NGC 4449, Wilcots \& Thurow (2001) found velocity differences
between the HI and \ha\ of order 50--100 \kms. These large differences
were associated with diffuse \ha\ in low column density HI gas,
interpretted as champagne flows away from giant \HII\ regions and
into holes in the ISM.

\subsection{Stars}

The \vhelio\ measured from the absorption lines 
are plotted against position along the slit in Figure \ref{fighelion4449abs}.
In the figure we show the results from measuring the 
absorption spectra summed in 10.65\arcsec\ steps
along the slit. This corresponds to intervals of 
200 pc in NGC 4449.
The stellar rotation velocities are also plotted with the neutral
and ionized gas in Figure \ref{fighin4449em}. 
We have not detected
rotation in NGC 4449 at any of the four position angles
to a radius of 65\arcsec\ (1.2 kpc).
The solid line in Figure \ref{fighelion4449abs} 
is the central velocity of 214 \kms.
The dashed line is the fit to the ionized gas rotation at
PA 46\arcdeg, and it is clear that we can rule out a stellar
rotation comparable to that of the ionized gas. We set an upper limit on
the rotation of the stars of (3/$\sin i$) \kms\ kpc$^{-1}$.


\section{Summary of observational results}

In NGC 2366 we have measured the rotation of the ionized gas
to 114\arcsec\ (1.9 kpc).
Rotation with a major axis PA of 30\arcdeg, that of the optical
galaxy, or of 45\arcdeg, that of the HI kinematics, fit the observations
equally well. 
The rotation gradient is linear (can be fit with a straight line) 
with a gradient of 
($30\pm2/\sin i$) \kms\ kpc$^{-1}$.
Peculiar velocities southwest of the galaxy center, where
there are two supergiant HII regions, mask the rotation pattern
there.

In NGC 1156 the ionized gas shows a rotation gradient that can be
fit with a straight line to 42\arcsec\ (1.6 kpc). The major axis
is 84\arcdeg---129\arcdeg.
The HI kinematics have a PA of order 130\arcdeg\
in the region the \ha\ is being measured, but a PA of order 83\arcdeg\ 
overall (Swaters 1999).
However, the optical major axis has a PA of 39\arcdeg,
so the morphological and kinematic axes are misaligned.
The gradient of the ionized gas is ($13\pm1/\sin i$) \kms\ kpc$^{-1}$.
Rotation of the stars is weak, but
linear to 43\arcsec\ (1.6 kpc).
The most probable kinematic major axis of the stars is also 84\arcdeg.
The gradient is ($5\pm1/\sin i$) \kms\ kpc$^{-1}$.

In NGC 4449 we found that the ionized gas rotates in a manner consistent 
with a major axis of 46\arcdeg, the same as the morphological
major axis of the inner galaxy.
The gradient is ($12\pm1/\sin i$) \kms\ kpc$^{-1}$ to 110\arcsec\ (2.1 kpc).
The stars, on the other hand,
show no ordered rotation, with an upper limit of 
(3/$\sin i$) \kms\ kpc$^{-1}$
to 65\arcsec\ (1.2 kpc).

In all three galaxies we have found that the ionized and neutral gas 
velocities are similar,
but often with important differences.
First, the rise of the
rotation speeds with radius is steeper in the ionized gas than
in the HI. This is easily understood as due to the large difference
in resolution of the two data sets. The HI observations typically were 
made with a beam-size of order 30\arcsec\ while the
optical spectra used a slit of 3\arcsec. This beam-smearing
effect can make the rotation curve deduced from the HI 
shallower (see, for example, 
Bosma 1981; Rubin \et\ 1989; 
Blais-Ouellette \et\ 1999;
Swaters, Madore, \& Trewhella 2000).

Second, the major axes inferred from the kinematics of the different
galactic components are
not always the same, nor are they the same as the major axis of the
optical morphology.
The various PAs are collected in Table \ref{tabpa}.
In NGC 1156 the various gas and stellar kinematical axes are most
likely the same, but this axis differs from the morphological major axis.
In NGC 2366 the morphological and HI kinematical axes are different,
and the ionized gas is equally consistent with either of these axes.
In NGC 4449 the kinematical
major axis of the ionized gas is coincident with the morphological
major axis of the inner optical galaxy but differs from the HI
kinematic axis.

In all three galaxies we can identify regions in which the ionized
gas rotation velocity gradients can be approximated by a straight line. 
However, in many regions the true rotation is disguised by large superposed
non-circular motions. 
The velocity gradient of the stars in NGC 1156 is also linear
over the region of the galaxy that we measured rotation velocities.
The gradients in the stars are lower than those in the ionized gas
in both galaxies.

\section{Discussion}

\subsection{NGC 1156}

\subsubsection{Misalignment of axes}

In NGC 1156 the stars and gas, both ionized and neutral, appear
to rotate about the same axis. A line of nodes of about 84\arcdeg\
is consistent with all three components of the galaxy.
However, this axis is different by about 45\arcdeg\ from the 
morphological major axis.

What does it mean when the line of nodes is misaligned
with the morphological major axis?
NGC 1156 has a rectangular appearance over much of the optical system.
Although there is no twisting of isophotes or other tell-tale signs,
this rectangular appearance could indicate the presence of a bar potential
(Athanassoula \et\ 1990).
In modeling a disk galaxy, one usually assumes 
a plane circular thin disk which is inclined to the line of
sight at a fixed angle. In a barred galaxy in which the bar
extends most of the radius, the isophotes are not circular
in the principle plane, the disk is not thin,
and the orbits may precess, so the 
expectation in a well-behaved
galaxy that the line
of nodes and morphological axis are the same no longer applies.
Identifying bar structures in irregular galaxies 
is complicated by the lack
of symmetry provided by
spiral arms coming out of the nucleus or bar (see discussion by
Roye \& Hunter [2000] and
Hunter \et\ [2001]).
Furthermore, the bar in NGC 1156 is a much larger
fraction of the optical galaxy than is the case in spirals, even
to the point where the bar is most of the optical galaxy.
In spirals, on the other hand, the bar length 
is typically $\leq$0.3 of the size of the stellar disk
(Elmegreen \& Elmegreen 1985). 
However, bar structures are very stable and they can grow on timescales
less than a Hubble time. So, while such large bars are not seen
in more luminous galaxies, it is reasonable that a rectangular galaxy
like NGC 1156 is dominated by a bar.

\subsubsection{Low stellar velocity gradient}

The apparent low velocity gradient in the stars compared to that of the 
ionized gas in NGC 1156 is also striking. The gradient in the stars is about 
3 times lower than that of the ionized gas. However, the uncertainties
in the stellar rotation velocities are large enough that we cannot
rule out the possibility that the stellar gradient is nearly the
same as that of the ionized gas.
In Figure \ref{fighelion1156abs} the ionized gas velocity gradient
is superposed on the stellar rotation velocities. In all but PA 84\arcdeg\
the ionized gas velocity gradient fits within the error bars of
the stellar points.
By the same argument, of course, we cannot rule out the possibility that 
the stellar velocity gradient is lower, even zero.

If the velocity gradient of the stars {\it is} lower than that of
the gas, 
it is possible that more of the kinetic energy of the stars
is in random motions rather than ordered rotation compared to the gas.
The rotation velocity of the ionized gas
is ($15.9\pm1.3/\sin i$) \kms\ at a radius of 31.3\arcsec\ where we measure
a rotation velocity for the stars of ($5.7\pm1.8/\sin i$) \kms.
If the galaxy is a disk with the ionized gas defining
the circular rotation speed, we can estimate the azimuthal velocity dispersion 
of the stars from $\sigma_{\phi}^2 \sim v_{gas}^2 - v_{stars}^2$
(see discussion in \S4.2.1(a) of Binney \& Tremaine 1987). 
This would yield a stellar velocity dispersion
of order (14.8/$\sin i$) \kms\ 
to compensate for the lower observed
rotation velocity, assuming the stars and gas are in the same disk.
For an inclination of 51\arcdeg, determined for the HI velocity field
(Swaters 1999), $\sigma_{\phi}$ would be about 19 \kms, comparable
to the value predicted for the M$_B$ of the galaxy 
(Bottema 1993, Swaters 1999, van der Marel \et\ 2002).

However, there is an inconsistency in this argument.
Although the absorption features that we
targeted are due to F---K-type stars, these may be dominated
by stars less than about 1 Gyr of age. The high surface brightness
region that we observed is blue and the V-band is dominated by
stars $\leq$1 Gyr old. 
It would be hard to impart a large level of randomization to
the velocities of the stars 
on this timescale. 
Also, there is no obvious structure at the center of the
galaxy to scatter the stars.

\subsection{NGC 4449}

\subsubsection{No measureable rotation of the stars}

NGC 4449 was known to be peculiar. 
Large streamers of HI that encircle the galaxy indicate
that another galaxy has interacted with NGC 4449 sometime
in the past (Hunter \et\ 1998). Furthermore,
the HI gas in the inner 2\arcmin\ of the galaxy appears to be
counter-rotating with
respect to the gas at larger radii, a clear signature of
an interaction.
The stars do not correspond to either gas system: {\it we detected
no ordered rotation} to $<3/\sin i$ \kms\ kpc$^{-1}$ to 1.2 kpc.
As for NGC 1156, the uncertainties in the stellar rotation velocities
are relatively high, but
the maximum allowed velocity gradient 
is still 4 times lower than the gradient observed
in the ionized gas. As Figure
\ref{fighelion4449abs} shows,
it is unlikely
that the stars are rotating at the same rate 
as the gas. 

What does it mean then when the stars do not rotate but the gas does?
One possibility is that 
the stars, in contrast to the gas, lie in a disk that is
nearly face-on.
An inclination of 45\arcdeg\ is
inferred from $b/a$ in the outer isophotes
of the galaxy under the assumption that the intrinsic axis ratio
is 0.3. However, NGC 4449, like NGC 1156, is rectangular in shape,
and there is a twisting of the V-band isophotes from
the central rectangle to the outer galaxy. Hunter \et\ (1999)
interpret these characteristics as a bar structure that 
has a length that is 90\% of D$_{25}$.
If the galaxy is dominated by a bar, the observed $b/a$ is
not a valid indicator of the inclination of the galaxy.

\subsubsection{A model for the gas kinematics}

If the stars are in a disk seen nearly face-on, 
how would we understand the kinematics
of the gas?
One explanation 
is that the HI disk is highly warped (Swaters 1999). 
We explored this possibility
by constructing a model for the gas in which the gas is
in orbits of constant inclination, but with precession-induced
twisting of the line of nodes. 

We used fits to the HI velocity field of NGC 4449 to guide the models.
Hunter \et (1998, 1999) have presented such fits, but,
because of the complex HI velocity field in this galaxy,
we have now calculated
additional fits to remind ourselves of the degree of uncertainty.
We fit the data from 15\arcsec\ to 2\arcmin\ in summed annular steps 
of 15\arcsec\ and from 135\arcsec\ to 12\arcmin\ in steps of 45\arcsec. 
Although we explored a variety of parameters, 
in the final fit we fixed the center as the optical center of the galaxy.
First, we allowed the inclination, 
PA, and rotation speed to vary with each annulus. 
Then we fixed the inclination
and allowed only the PA and rotation speed to vary.
We found
an inclination of 71$\pm$9\arcdeg\ and a PA that varied
between 80\arcdeg\ and 105\arcdeg. 
These values are close to those of Hunter \et\ (1999) for the inner
galaxy---an inclination
of 77\arcdeg\ at a position angle of 72\arcdeg. 
Thus, the present model uses
a line of nodes of 80\arcdeg\ in the central 2\arcmin.
For the outer galaxy, we found an inclination of 66$\pm$17\arcdeg\
with the position angle settling down to 253$\pm$4\arcdeg\ beyond
6\arcmin. Hunter \et\ (1998) had found an inclination of 60\arcdeg\
and a position angle of 230$\pm$17\arcdeg. 
The models of Theis \& Kohle (2001) that successfully reproduce the extended gas
streamers constrain the outer gas disk to an inclination of 50--70\arcdeg.
Our model uses
a line of nodes of $-110$\arcdeg\ ($=250$\arcdeg) in the outer regions
(r$>$5\arcmin).

The model that we present here
extends to a radius of 10\arcmin, has a disk inclination of 68\arcdeg,
and uses a systemic velocity of 214 \kms.
The maximum circular velocity is 80 \kms, and the velocity
ramps linearly from 0 to 80 \kms\ between r$=$0 and r$=$2.5\arcmin.
The line of nodes in the inner 2\arcmin\ is 80\arcdeg\ and 
decreases linearly with radius to
$-$110\arcdeg\ at a radius of 5\arcmin. 
A three dimensional representation of the model orbits is shown
in Figure \ref{figmod}.
Physically, this model corresponds to a disk of gas 
in tilted orbits which precess
around the minor axis of an oblate gravitational potential.
Precession periods are shorter at smaller radii, so the line of
nodes regresses, and the gas disk twists.
The line of nodes precesses retrograde to rotation in an oblate 
potential, and since precession periods are shorter at smaller radii,
the line of nodes develops a leading twist with radius.
Such orbits can be stable until cloud-cloud collisions produce dissipation,
so they can last a long time. 

We compare the model to the observations in Figure \ref{figmodonhi}.
Here we reproduce the HI position-velocity plots 
from Figure \ref{fighin4449em} and superpose simulated observations
of the model at 15\arcsec\ intervals at our four PAs.
One can see that the model reproduces well the observed HI 
rotation velocities including the counter-rotation in the inner
galaxy. 
It predicts broad or multiple profiles in the HI in places:
particularly at $\pm$2\arcmin--2.5\arcmin\ along PA 46\arcdeg\ and
$\pm$1.5\arcmin--2\arcmin\ along PA 91\arcdeg.
Plots of HI profiles in the inner galaxy do show regions with broad
and double-peaked profiles (broad profiles are easily seen
in the position-velocity plots of Figure \ref{figmodonhi}),
although not exclusively at the regions
predicted by the models, which
do not take account of beam-smearing in the HI observations.
We conclude that this simple model is a plausible
representation of the orbits of the gas in NGC 4449.

Such tilted orbits would result from 
tipping gas in at an angle to the stellar disk.
Most probably the gas was acquired from a passing or merging object.
Since there are other signs that NGC 4449 was involved in 
a past interaction, we can understand the gas in these peculiar
orbits as resulting from that interaction or a second event.
The models of Theis \& Kohle (2001) that reproduce the extended streamers
as resulting from 
an interaction with DDO 125 involve no mass transfer and do not
reproduce the apparent counter-rotating gas system in the center.
If the acquired gas was simply gas in 
NGC 4449 pulled out by a passing galaxy and
now settling back in, we would expect stars to have been pulled
out into tilted orbits also, which we
do not see.
Thus, an accretion event is required.

Hunter et al. (1999) had argued
that NGC 4449 before the interaction
had to have been a late-type galaxy and, therefore,
likely gas-rich. 
This argument was based on the ensemble of characteristics including
small M$_B$, small optical scale length, lack of a bulge,
small rotation speed, current high gas content, and low metallicity.
If NGC 4449 was relatively gas-rich to begin with, 
the acquired gas is falling onto other
gas that was already there, presumably in the same orbits as the stars.
The original disk gas could have lost angular momentum in the
encounter and been pushed further in, and is
now a small inner nearly-face-on disk. However, the HI observations do
not allow such a face-on disk of original material to be very massive;
the entire HI mass within r$<$2\arcmin\ is only $3\times10^8$ M\solar\
(Hunter \et\ 1999), and the center of the galaxy represents 
a minimum in the HI distribution.

In the models, the twisting of the line of nodes of the 
gas takes place
from a radius of 2\arcmin\ to a radius of 5\arcmin.
The region 2\arcmin$<$r$<$3\arcmin\ is a special place in NGC 4449
in several other ways as well.
First, the morphological major axis changes PA in this 
region.
The PA of the morphological major axis rotates from 
46\arcdeg\ for the rectangular inner galaxy (r$<$2.2\arcmin) 
to 64\arcdeg\ at r$>$3.1\arcmin. If the rectangular structure of
the galaxy is a bar, this region represents the end of the bar potential.
Second, there are large HI complexes located at a radius of 
1.5\arcmin--2\arcmin.
These HI complexes are part of a large HI ring that surrounds the
optical galaxy.
Gas could be collecting there both because it is trapped near
the end of the bar,
and because infalling gas is piling on with an increased number
of cloud-cloud collisions. The cloud-cloud collisions would further
account for the star formation that is currently seen there.

Sung \et\ (1998) have argued that irregular galaxies are triaxial in shape.
In NGC 4449, the twisting of the isophotes (de Zeeuw 1984) and
the rectangular shape (Binney \& Tremaine 1987)
could be consistent with a triaxial shape with stars 
on box and loop orbits. However, few elliptical galaxies, which 
as a class include triaxial systems, are
as boxy as NGC 4449 (See Peleier \et\ 199; Figure 2 of Ryden \et\ 1999;
Figure 6.1 of Sparke \& Gallagher 2000; but also see Figure 2 of
Jarvis 1987), 
although bar structures can be (Athanassoula \et\ 1990).
Since we can fit the observations of NGC 4449 
with a reasonable model involving gas in a warped plane relative
to the stars, and have an obvious mechanism for putting gas into these orbits,
we do not embrace more complicated explanations.


\subsection{Summary of interpretation}

We measure a weak velocity gradient in the stellar disk of NGC 1156 
and no rotation of the stars in NGC 4449.
We find that we can understand the stellar and gas kinematics of NGC 1156
and NGC 4449 as disk systems dominated by bar potentials.
In both galaxies the bar contains a large fraction
of the light of the optical galaxy.

The low stellar velocity gradient of the stars in NGC 1156 is
consistent with the velocity gradient of the gas within the
observational uncertainties.
A consistent and satisfying picture of NGC 1156
results if the stellar velocity gradient is close to that
of the ionized gas and all components are in circular rotation
about a common axis. 
The misalignment of the kinematical axis with the morphological
axis is the result of a large bar dominating the disk luminosity.

In NGC 4449, where the stars rotate much more slowly than the gas,
we can qualitatively explain the observations 
by assuming that
the stars are in a disk seen nearly face-on, while the 
gas lies 
in a tilted disk with precession-induced twisting of the
line of nodes. Most likely, the gas in this tilted disk
was acquired from an interaction or possibly a merger, consistent
with other signs of a past interaction in the galaxy.

Further clarification and refinement
of our picture of the stellar kinematics
in both NGC 1156 and NGC 4449 will have to await a measurement of the stellar
velocity dispersions in these systems.

\acknowledgments

We wish to thank Phil Massey, Skip Andree, and other members of
the KPNO staff for help in making the 4 m telescope and RC spectrograph
perform so well for us. 
Support for this work was provided to DAH by the Lowell Research Fund
and in part by the National Science Foundation
through grant AST-9616940 to DAH and grant AST-0098419 to LSS.

\clearpage
 
\begin{deluxetable}{rccccrr}
\tablecaption{Characteristics of the galaxies. \label{tabgal}}
\tablewidth{0pt}
\tablehead{
\colhead{} & \colhead{D} & \colhead{} 
& \colhead{} & \colhead{D$_{25}$} 
& \colhead{RA} 
& \colhead{DEC} \\
\colhead{Galaxy} & \colhead{(Mpc)} & \colhead{E(B$-$V)$_f$} 
& \colhead{M$_{B,0}$} & \colhead{(kpc)} 
& \colhead{(2000)} 
& \colhead{(2000)} \\
\colhead{(1)} & \colhead{(2)} & \colhead{(3)} 
& \colhead{(4)} & \colhead{(5)} & \colhead{(6)} & \colhead{(7)}
} 
\startdata
NGC 1156 & 7.8 & 0.165 & $-$18.0 & 7.5 & 2 59 42.3 & 25 14 15 \\
NGC 2366 & 3.4 & 0.043 & $-$16.5 & 5.2 & 7 28 52.3 & 69 12 30 \\
NGC 4449 & 3.9 & 0.000 & $-$18.2 & 7.2 & 12 28 11.0 & 44 05 36 \\
\enddata
\tablecomments{Column notes:
(2) For NGC 1156 and NGC 4449 distances are determined from 
radial velocities corrected to the Galactic Standard of Rest (RC3)
and a Hubble constant of 65 \kms\ Mpc$^{-1}$.
For NGC 2366 the distance comes from cepheid light curves
(Tolstoy \et\ 1995).
(3) Foreground reddening from Burstein \& Heiles (1984).
We use the extinction curve of Cardelli, Clayton, \& Mathis (1989)
to correct for extinction.
(4) Integrated M$_B$ corrected for extinction.
Data for NGC 1156 from RC3; data for NGC 2366 from
Hunter \et\ (2001); data for NGC 4449 from
Hunter \et\ (1999).
(5) The diameter of the galaxy measured to an extinction-corrected
B-band surface
brightness of 25 magnitudes arcsec$^{-2}$
(NGC 1156: RC3; NGC 2366:
Hunter \et\ (2001); NGC 4449:
Hunter \et\ (1999).
(6--7) The center of the galaxy determined from V-band images
and used to place the center of the slit. 
Units of RA are hours, minutes, seconds; 
units of DEC are degrees, arcminutes, and arcseconds.
}
\end{deluxetable}

\clearpage
 
\begin{deluxetable}{clcc}
\tablecaption{Template stars observed on each night. \label{tabstars}}
\tablewidth{0pt}
\tablehead{
\colhead{Night} & \colhead{Star} & \colhead{Type} 
& \colhead{\vhelio\tablenotemark{a} (\protect\kms)}
} 
\startdata
1 & HD 10380   & K3III    & \nodata \\
  & HD 41636   & G9III    & \nodata \\
  & HD 132142  & K1V      & \nodata \\
  & HD 145328  & K0III-IV & \nodata \\
2 & HD 9138    & K4III    & $+$35.4 \\
  & HD 5211    & F4III    & \nodata \\
3 & HD 4338    & K3III    & $-$28.3 \\
  & HD 107328  & K0.5IIIb & $+$35.7 \\
4 & HD 1069    & K2I      & \nodata \\
  & HD 1400    & K5I      & \nodata \\
  & HD 18884   & M1.5III  & $-$25.8 \\
\enddata
\tablenotetext{a}{Radial velocity standards are taken from the
United States Nautical Almanac (2000).
}
\end{deluxetable}

\clearpage
 
\begin{deluxetable}{rccccccc}
\tablecaption{Central velocities V$_{sys}$ (km s$^{-1}$). \label{tabvc}}
\tablewidth{0pt}
\tablehead{
\colhead{} & \colhead{} & \colhead{}
& \colhead{} & \colhead{$\Delta$V$_{sys}$} & \colhead{$\Delta$V$_{sys}$}
& \colhead{} 
& \colhead{} \\
\colhead{Galaxy} & \colhead{Slit PA} & \colhead{Ionized Gas}
& \colhead{Stars} & \colhead{Ionized Gas} & \colhead{Stars}
& \colhead{Average} 
& \colhead{HI\tablenotemark{a}}
} 
\startdata
NGC 1156 &  39 & 383.4$\pm$2.1 & 379.7$\pm$0.4 & $-$2.2 & $+$1.5 & 381.2$\pm$2.1 & 371 \\
         &  84 & \nodata       & \nodata       & $+$0.0 & $+$7.9 &               &     \\
         & 129 & 382.7$\pm$1.4 & 371.5$\pm$1.8 & $-$1.5 & $-$0.3  &               &     \\
         & 174 & \nodata       & 378.5$\pm$0.7 & $-$3.3 & $+$2.7 &               &     \\
NGC 2366 &  30 &  95.2$\pm$2.3 & \nodata       & $+$0.8 & \nodata & 96.0$\pm$0.7 & 98--104 \\
         &  27 &  96.4$\pm$2.3 & \nodata       & $-$0.4 & \nodata &              &     \\
         &  63 &  \nodata      & \nodata       & $-$0.5 & \nodata &              &     \\
         & 120 &  96.5$\pm$0.9 & \nodata       & $-$1.8 & \nodata &              &     \\
NGC 4449 &   1 &  \nodata      & 212.2$\pm$1.4 & $-$4.8 & $+$1.8 & 214.0$\pm$2.5 & 208--214 \\
         &  46 & 218.2$\pm$1.0 & 212.4$\pm$1.0 & $-$4.1 & $+$1.7 &               &     \\
         &  91 & \nodata       & 213.1$\pm$1.5 & $-$2.8 & $+$1.0 &               &     \\
         & 139 & \nodata       & 214.4$\pm$1.7 & $-$8.5 & $-$0.3 &               &     \\
\enddata
\tablenotetext{a}{References for HI: NGC 1156---Swaters 1999; 
NGC 2366---Hunter et al.\ 2001 and references therein;
NGC 4449---Hunter et al.\ 1998, 1999.}
\end{deluxetable}

\clearpage
 
\begin{deluxetable}{rrc}
\tablecaption{Position angles. \label{tabpa}}
\tablewidth{0pt}
\tablehead{
\colhead{} & \colhead{} & \colhead{PA} \\
\colhead{Galaxy} & \colhead{Data} & \colhead{(deg)} \\
} 
\startdata
NGC 1156 & V-band        &  39 \\ 
         & HI kinematics (overall) &  83 \\ 
         & HI kinematics (inner)   & 130 \\
         & Ionized gas kinematics & 84---129 \\ 
         & Stellar kinematics     & 84       \\ 
NGC 2366 & V-band        &  32 \\ 
         & HI kinematics &  46 \\ 
         & Ionized gas kinematics & 30, 45 \\ 
NGC 4449 & V-band (r$<$2.2\arcmin) & 46 \\ 
         & V-band (r$>$3.1\arcmin) & 64 \\ 
         & HI kinematics (r$<$2\arcmin) & 80---105 \\ 
         & HI kinematics (6\arcmin$<$r$<$12\arcmin) & 73 \\ 
         & Ionized gas kinematics & 46 \\ 
\enddata
\end{deluxetable}

\clearpage

\figcaption{False-color representation of the logarithm of the 
V-band and \protect\ha\ images of our galaxies. 
We show the logarithm of the V-band image in order to allow comparison
of different surface brightness levels.
In the V-band images a line is drawn along the
morphological major axis. All four PAs observed in each galaxy
are shown in the \protect\ha\ images. 
The circle marks the center of the galaxy as given in Table \protect\ref{tabgal}.
The ticks along the lines that show the PAs mark the 
extent over which we measured
the absorption features (in the V-band images) and emission features
(in the \protect\ha\ images).
The V-band and \protect\ha\ images for NGC 2366 come from Hunter \protect\et\
(2001) and for NGC 4449 from Hunter et al.\ (1999).
The V-band image of NGC 1156 was kindly obtained for us by P.\ Massey with
a Tektronic 2048$\times$2048 CCD on the KPNO 4 m telescope. 
The \protect\ha\ image of NGC 1156 was obtained with the Perkins 1.8 m telescope
at Lowell Observatory in 1995, using a TI CCD and a 32 \protect\AA\ FWHM
interference filter and 95 \AA\ off-band filter.
\label{figv}}

\figcaption{Line-of-sight 
heliocentric radial velocities \protect\vhelio\ measured
from the emission line [OIII]$\lambda$5007 in the spectra of NGC 2366.
Position along the slit is measured from the center of the galaxy,
and positive numbers refer to the side of the slit 
that is in the direction of
the given PA.
Model 1 (major axis of
30\protect\arcdeg, solid line), and 
Model 2 
(major axis of 45\protect\arcdeg, dashed line) are
discussed in \S \protect\ref{sec2366}.
The lines are found from linear fits over the radial extent of the 
lines shown.
The giant HII regions NGC 2363 and NGC 2366-III (Drissen et al.\ 2000)
are marked along with a region of filaments near NGC 2363
(Hunter \et\ 1993).
\label{fighelion2366}}

\figcaption{Line-of-sight 
heliocentric radial velocities of the ionized gas are plotted
on contours of the neutral HI gas in NGC 2366. 
The HI position-velocity plots are taken from data presented by
Hunter \protect\et\ (2001).
The HI was summed over 35\protect\arcsec, approximately one beam-width,
and contours are 0.88, 3.5, 11.8, and 35.3$\times10^{20}$ \protect\coldens.
There could be a small offset in V$_{sys}$ between the optical and HI
(see Table \protect\ref{tabvc}).
\label{fighin2366em}}

\figcaption{Emission-line [OIII]$\lambda$5007
line-of-sight heliocentric velocities 
in NGC 2366 in PA 27\arcdeg, 30\arcdeg, and
63\arcdeg\ are superposed over the limited region 
of $+$125\protect\arcsec\ $<$ r $<$ $-$50\protect\arcsec.
The solid line is the fit to the data and is designated Model 2 
(kinematical major axis of
45\protect\arcdeg).
\label{fign2366mod2}}

\figcaption{Rotation velocity
V$\rm _{rot}$ plotted against distance from the center of the galaxy
in the plane of the galaxy for NGC 2366.
The points are the data at PA 27\protect\arcdeg\ and 30\protect\arcdeg\
converted to rotation velocity using a
position angle of 30\protect\arcdeg\ (Model 1) and an inclination
of 72\protect\arcdeg.
The solid line is the Model 1 fit to the data (major axis PA=30\protect\arcdeg),
and the tilted dashed line is the Model 2 fit to the data (major axis
PA=45\protect\arcdeg).
The lengths of the lines indicate the regions over which the fits were made.
The long-dashed, curved line is the HI rotation curve
(Hunter \protect\et\ 2001).
The short-dashed horizontal
line at a rotation speed of 0 is intended to assist the eye.
\label{figrot2366}}

\figcaption{Line-of-sight 
heliocentric radial velocities \protect\vhelio\ measured
from the emission lines of the ionized gas in each spectrum
of NGC 1156.
The position angle of each spectrum is given; the top left panel is
the morphological major axis.
Position along the slit is measured from the center of the galaxy,
seen as the intersection of all the slits in Figure \protect\ref{figv},
and positive numbers refer to the direction of the given PA.
The solid line is a fit to the velocities in PA$=$129\arcdeg\ 
from r$=+$29.9\arcsec\ to r$=-$36.6\arcsec.
The length of the line indicates the region over which the fit
was made.
The solid lines in the other PAs are the fit at
PA$=$129 transformed to the given PA in the plane of the sky
assuming a kinematical major axis of 129\arcdeg\ and an 
inclination angle of 35\arcdeg. 
The dashed line is the fit to the data in PA 84\arcdeg\ 
from r$=+$42.2\arcsec\ to r$=-$19.2\arcsec\ 
and the 
transformation of that fit to the other PAs in the plane of the sky
assuming the kinematic major axis is along PA 84\arcdeg.
The place where the PA 39\arcdeg\ slit cuts through the
large HII complex to the southwest is marked.
\label{fighelion1156em}}

\figcaption{Line-of-sight heliocentric radial velocities \protect\vhelio\
of the ionized gas and stars are compared
to that of the neutral HI gas (contours) in NGC 1156. 
The HI position-velocity plot is taken from data presented by
Swaters (1999).
The HI was summed over 30\protect\arcsec, approximately one beam-width,
and contours are 1.0, 2.1, 3.2, and 4.3$\times10^{20}$ \protect\coldens.
There could be a small offset in V$_{sys}$ between the optical and HI
(see Table \protect\ref{tabvc}).
\label{fighin1156em}}

\figcaption{Line-of-sight 
heliocentric radial velocities \protect\vhelio\ measured
from the stellar absorption lines in each spectrum
of NGC 1156.
The absorption measurements are 10.65\protect\arcsec\ sums stepped
along the slit.
The position angle of each spectrum is given; the top left panel is
the morphological major axis.
Position along the slit is measured from the center of the galaxy,
seen as the intersection of all the slits in Figure \protect\ref{figv},
and positive numbers refer to the direction of the given PA.
The solid lines are the rotation curve determined from a least-squares
fit to the points in PA$=$84\arcdeg\ from r$=+43.3$\arcsec\ (1.6 kpc) to
r$=-41.9$\arcsec\ (the length of the line) and transformed to the other
position angles in the plane of the sky, assuming an inclination of 
35\arcdeg.
The dashed line is the fit to the rotation of the ionized gas
(the dashed line from Figure \protect\ref{fighelion1156em}), 
plotted here for comparison.
\label{fighelion1156abs}}

\figcaption{Rotation velocity 
V$\rm _{rot}$ plotted against distance from the center of the galaxy
in the plane of the galaxy for the stars in NGC 1156.
The solid horizontal
line at a rotation speed of 0 is intended to assist the eye.
The solid line is the fit to PA 84\arcdeg\ 
transformed to the plane of the galaxy using an inclination of 35\arcdeg.
The length of the line indicates the region over which the fit 
was performed.
The dashed line is the rotation curve determined from the HI
velocity field (Swaters 1999) reflected about r$=$0; 
from the fit to the HI velocity field the PA is 83\arcdeg\ and 
the inclination is 51\arcdeg.
The error bar in the lower right corner gives the median uncertainty
in the velocities.
\label{figrot1156}}

\figcaption{Line-of-sight 
heliocentric radial velocities \protect\vhelio\ measured
from the emission lines of the ionized
gas in the spectra
of NGC 4449.
The position angle of each spectrum is given; the top left panel is
the morphological major axis for r$<$2.2\arcmin\ determined 
from the V-band image.
Position along the slit is measured from the center of the galaxy,
seen as the intersection of all the slits in Figure \protect\ref{figv},
and positive numbers refer to the direction of the given PA.
The solid line in PA$=$46\protect\arcdeg\ is a least-squares fit
to the portion of the data with a linear rotation gradient
(excluding end points) from r$=+110.1$\arcsec\ to r$=-62.7$\arcsec.
The line is shown only as long as the region fit.
The solid lines in the other PAs are transformations
of this fit to the given PA in the plane of the sky
under the assumption of an inclination
angle of 45\protect\arcdeg.
Peculiar velocities that are concident with obvious shells and filaments
are marked; features identified as ``HHG'' are catalogued in
Hunter, Hawley, \& Gallagher (1993).
\label{fighelion4449em}}

\figcaption{Line-of-sight 
heliocentric radial velocities of the ionized gas and stars 
are compared
to that of the HI in NGC 4449. 
The HI position-velocity plots are taken from data presented by
Hunter \protect\et\ (1999).
The HI was summed over 10\protect\arcsec, approximately one beam-width,
and contours are 3.75, 11.25, and 18.75$\times10^{20}$ \protect\coldens.
There could be a small offset in V$_{sys}$ between the optical and HI
(see Table \protect\ref{tabvc}).
\label{fighin4449em}}

\figcaption{Line-of-sight 
heliocentric radial velocities \protect\vhelio\ measured
from the stellar absorption lines 
in each spectrum
of NGC 4449.
The absorption measurements are steps along the slit 
summed over 10.65\protect\arcsec.
The position angle of each spectrum is given; the top left panel is
the morphological major axis determined from the V-band image.
Position along the slit is measured from the center of the galaxy,
seen as the intersection of all the slits in Figure \protect\ref{figv},
and positive numbers refer to the direction of the given PA.
The solid lines are the central velocity.
The dashed line is the fit to the rotation velocities of the ionized gas
at PA 46\arcdeg, as shown in
Figure \protect\ref{fighelion4449em}, for comparison.
\label{fighelion4449abs}}

\figcaption{Spatial representation of the orbits of the gas
in the model of NGC 4449: a disk of gas in which tilted
orbits precess around the minor axis in an oblate spherical
potential. At smaller radii where the precession periods 
are smaller, the line of nodes regresses from 80\protect\arcdeg\
(r$<$2\protect\arcmin) to $-110$\protect\arcdeg\ (r$>$5\protect\arcmin),
so the gas disk twists. See text for details.
{\it Top}: The three-dimensional model is shown in blue
and the projection on the plane of the sky is shown in red.
{\it Bottom}: The model is shown projected onto the plane of the sky.
The solid and dashed blue lines mark the position angles of
46\arcdeg\ and 136\arcdeg\ in the plane of the sky.
The line of nodes, shown as the solid black line, 
is in the plane of the sky in both panels.
The two panels are not at the same scale.
\label{figmod}}

\figcaption{Model of the gas orbits compared to the observed
HI velocities in NGC 4449.
The HI position-velocity plots are the same as in Figure 
\protect\ref{fighin4449em}. The model has been ``observed''
every 15\protect\arcsec\ along the same position angles as
the optical observations. The error bars indicate the 
velocity ``dispersions'' due to 
multiple velocities along the line of sight at those radii. 
\label{figmodonhi}}

\end{document}